\documentstyle[preprint,aps,pre,psfig,epsf]{revtex}
\newcommand{\beq}{\begin{equation}}
\newcommand{\eeq}{\end{equation}}
\newcommand{\bmath}{\begin{math}}
\newcommand{\emath}{\end{math}}
\newcommand{\bc}{\begin{center}}
\newcommand{\ec}{\end{center}}
\newcommand{\beqa}{\begin{eqnarray}}
\newcommand{\eeqa}{\end{eqnarray}}
\newcommand{\beqs}{\begin{eqnarray*}}
\newcommand{\eeqs}{\end{eqnarray*}}

\begin{document}
%\twocolumn[\hsize\textwidth\columnwidth\hsize\csname@twocolumnfalse\endcsname

\title{Learning and generation of long-range correlated sequences}

\author{A Priel 
\footnote{http://faculty.biu.ac.il/$\sim$priel} and  
I Kanter}
\address{Minerva Center and Department of Physics, 
        Bar-Ilan University, 52900 Ramat-Gan,
Israel}

\date{\today}
\maketitle

\begin{abstract}
We study the capability to learn and to generate long-range, power-law
correlated sequences by a fully connected asymmetric network.
The focus is set on the ability of neural networks to extract 
statistical features from a sequence. 
We demonstrate that the average power-law behavior is learnable, namely, 
the sequence generated by the trained network obeys the same statistical 
behavior. The interplay between a correlated weight matrix and the
sequence generated by such a network is explored. A weight matrix 
with a power-law correlation function along the vertical direction, 
gives rise to a sequence with a similar statistical behavior.
\end{abstract}
\vspace*{.2cm}
\vspace*{.3cm}

%]

\section{Introduction}

Real-life (temporal) sequences are characterized by a certain degree of 
correlation. It is known that a wide range of systems in nature displays 
long-range correlations, e.g., biological - DNA sequences and heartbeat 
intervals, natural - languages, etc., see \cite{bunde94,kan_kes95}.
Since long-range correlations can appear in many forms, we restrict the 
analysis to the case of power-law correlations in a random sequence, 
e.g., the correlation function for a 1D sequence $x_i$ 
is given by  
\beq\label{1d_cf}
C(l)=  ~\left\langle~ x_i ~ x_{i+l} ~\right\rangle~ \propto ~l^{- \gamma} 
\quad (l \rightarrow \infty~,~\gamma>0) \quad ,
\eeq
\noindent
where the angular brackets denote an average over the randomness.
This type of random sequence is also termed ``colored'' or correlated-noise. 
The general form of description allows us to investigate the capability of 
the network to capture statistical properties of a sequence. 

The theory of learning from examples by a neural network, and in particular 
on-line learning, has been developed almost exclusively for uncorrelated 
patterns, see \cite{riegler95,opper95}.
Though some particular cases of correlated patterns were treated, 
they were limited to simple spatial correlations within each pattern, 
or to temporal correlations of each input unit, e.g., \cite{tarkowski92}.
The case of long-range correlations is absent. Clearly,
the problem of extracting a feature from a correlated sequence whose length is
much larger than the network's size, cannot be treated under the same 
assumptions. Rather than dealing with the question of the
generalization  error 
(average over a distribution of patterns), we focus on the capability of the 
network to asymptotically capture the correlations within the sequence and 
its ability to generate a sequence with similar properties. 

As shown previously, the sequence generator (SGen), a continuous-valued
feed-forward network in which the next state vector is determined from past
output values, exhibits (quasi) periodic attractors in the stable regime,
regardless of the complexity of the weights, both in the case of a perceptron 
as well as multilayer SGen's, e.g., \cite{sgen95,martin98,liat98}; 
the unstable, chaotic regime is studied in \cite{pri99}.
Therefore, it is obvious that the perceptron-SGen, or its extension to 
multilayer networks, are not suitable candidates for learning and generating 
correlated noise. The natural way to overcome this limitation is to 
increase the complexity of the feedback in the architecture.
In this paper we study a fully connected, asymmetric network.
The updating rule for the network's state is either sequential or parallel,
namely, each unit is updated on its turn or all units are updated 
simultaneously. Unit $i$ $(i=1,\ldots,N)$ is updated as follows 
\beqa
S_i^{t+1} & = & \tanh \left( \beta \left[ \sum_{j=1}^{i-1} W_{ij} S_j^{t+1} + 
\sum_{j=i}^{N} W_{ij} S_j^t \right] \right) \label{updating_rule_seq} \\
S_i^{t+1} & = & \tanh \left( \beta \sum_{j=1}^{N} W_{ij} S_j^{t} \right)
\label{updating_rule_par}
\eeqa
\noindent
where eq.\ \ref{updating_rule_seq}(\ref{updating_rule_par})  
refers to the sequential(parallel) rule;
${\bf W}$ is an [NxN] weight matrix and $\beta$ is a gain parameter.
The network generates iteratively an infinite sequence $\{\sigma_m\}$ 
starting from an initial state, $\bf{S}^0$, as follows: 
\beq\label{gen_sequence}
\sigma_m=S_i^{t+1}~,~m=tN+i=1,2,\ldots \quad , 
\eeq
\noindent
where $ i=1,\ldots,N, ~~t=0,1,2 \ldots ~$.

Two complementary issues are discussed in this paper:
(a) Given a training sequence characterized by long-range correlations,
can we train a network in an on-line scheme to generate a sequence with the 
same asymptotic statistical properties? (b) The inverse problem,
is there an interplay between a network whose 
weight matrix follows a power-law correlation function, and the sequence
it generates?

It is important to stress that the model we investigate
is not proposed as a practical method for generating long-range correlated 
sequences, rather, it is motivated by the issues raised above.

In the next section we investigate the first question. 
An ``on-line'', gradient-based learning rule is applied 
where each example is presented to the network only once. 
The sequences that constitute the basis for the training patterns
and the correlated weight matrices, are generated using an algorithm for 
re-shaping the power spectrum of an uncorrelated sequence; the method has 
been developed for investigating various stochastic processes, 
see \cite{makse96}. 
In section \ref{sec_col_net} the inverse problem is analyzed. A method for
constructing the weight matrix is presented based on the findings obtained 
from section \ref{sec_gen_rule} regarding the correlation properties of the 
weights in trained networks. A simple analytical derivation support these
findings.

\section{Generalizing the rule of a colored sequence}
\label{sec_gen_rule}

Suppose a source generating sequences that obey eq.\ \ref{1d_cf}; the 
question we address in this section focuses on the possibility of learning the 
statistical properties of the source, and in particular the exponent 
$\gamma$ of the power-law correlation function. 
For a network defined by its weights ${\bf W}$, a gain $\beta$ and a 
nonlinear function $f$, the response of the ith unit, given the current state 
${\bf \xi}^{t}$, is 
\[
S^{t+1}_i = f({\bf \xi}^t,{\bf W}_i),
\]
\noindent
where ${\bf W}_i$ denotes the vector of weights connected to the ith unit.
The on-line learning algorithm minimizes a quadratic error function 
\beq
\epsilon_i({\bf \xi}^t, {\bf W}_i)=[S^{t+1}_i-\tau^{t+1}_i]^2/2 ~~, 
\eeq
\noindent
where $\tau^{t+1}_i$ is the desired response of the ith unit given the state 
${\bf \xi}^{t}$. The weights are updated according to 
\beq\label{weight_update}
{\bf W}^{t+1}_i = {\bf W}^{t}_i - {\eta \over N} \bigtriangledown_W 
\epsilon_i({\bf \xi}^t, {\bf W}^{t}_i) \quad ,
\eeq
\noindent
i.e., a gradient descent rule with a learning rate $\eta$
(similar results were obtained using the Hebbian learning rule).

The training patterns are defined as follows. 
Let $D_L=\{ x_1,x_2,\ldots,x_L \}$ be a 1D sequence obeying 
eq.\ \ref{1d_cf}.
A training pattern (the pair $({\bf \xi}^m, \tau^m_i)$ ~ $0<m\leq L-2N+1$, ~
$ i=1,\ldots,N $) is defined by
\beq\label{train_set}
\left\{
\begin{array}{ccl}
{\bf \xi}^m & = & (x_m, x_{m +1}, \ldots, x_{m +N-1})
\nonumber   \\
\tau^m_i & = & x_{m +N+i-1} 
\end{array}
\right. ~ ,
\eeq
\noindent
where each weight vector ${\bf W}_i$ is updated with the corresponding 
desired output $\tau^m_i$, and the same vector ${\bf \xi}^m$.
Updating all $N$ weight vectors for a given pattern, 
$({\bf \xi}^m, \tau^m_i ~i=1,\ldots,N)$, 
accounts for a single training cycle. The patterns for consecutive training  
cycles are achieved via sliding a window of size $N$ by one site along the 
sequence $D_L$; e.g., given a current pattern starting at site $m$ along the 
training sequence, ${\bf \xi}^m$ (eq.\ \ref{train_set}), 
the next pattern is ${\bf \xi}^{m+1} = (x_{m+1}, x_{m+2}, \ldots, x_{m+N})$. 
The training patterns can be obtained in a different scheme by
sliding the window $N$ sites (non-overlapping windows); i.e., 
\beq\label{train_set_par}
\left\{
\begin{array}{cclr}
{\bf \xi}^{m+1}& =& (x_{mN+1}, x_{mN+2}, \ldots, x_{mN+N}) 
\nonumber   \\
\tau_i^{m+1}&=&x_{(m+1)N+i} & m=0,1,\ldots
\end{array}
\right. ~ .
\eeq
\noindent
Results obtained in both schemes are similar, however, the length of
the sequence ($D_L$) used in the second scheme (to obtain the same results)
is about $N$ times larger. 

Let us now describe our numerical investigation.  
An ensemble of long sequences, $D_L$ (of size $L \gg N $), 
that obey eq.\ \ref{1d_cf} with a given $\gamma$ is generated. 
A randomly chosen network is trained using a part of the sequence. 
Taking the last pattern from the training process as an initial state,
the trained network is used to generate iteratively a long sequence 
$\{\sigma_m\}$, of size $M N$ with $M=10$, following the sequential rule, 
eq.\ \ref{updating_rule_seq} and eq.\ \ref{gen_sequence}.
The correlation function of this sequence can be calculated in the two
following ways: spatial and temporal. 
The spatial correlation is obtained by averaging the correlation function, 
calculated on ($M$) sequences of size $N$ after updating all ($N$) units, 
over the $M$ iteration cycles, whereas the temporal 
correlation is simply the correlation function of the long sequence, i.e., 
\beq\label{cor_spat_temp}
\begin{array}{ccl}
C_{spat}(l) & = & 1 / M \sum_{j=0}^{M-1} \left[1/N 
\sum_{i=1}^N \sigma_{\tiny jN+i} ~\sigma_{\tiny jN+(i+l\bmod N)} \right] \\ 
C_{temp}(l) & = & (M N)^{-1} \sum_{i=1}^{M N} \sigma_i  \sigma_{i+l} ~ ,
\end{array}
\eeq
\noindent
where we take periodic boundary conditions. Next, the same weight
matrix is further 
trained and after each $\alpha N$ patterns ($\alpha=10$) , the same process 
of ``generating a sequence and calculating its correlation function'', 
is repeated for a better statistical estimation. 
We found that both definitions of the correlation function yield 
similar results, therefore, in the sequel we omit the subscript from eq.\ 
\ref{cor_spat_temp} and refer to the temporal correlation function
\beq\label{cf_def}
C(l)=C_{temp}(l) \quad .
\eeq
\noindent
Note that the range of correlations is bounded by the number of degrees of 
freedom $\{S_i\}_{i=1}^N$, as in \cite{makse96}; therefore, the correlation 
function is calculated in the range $0<l<N/2$ (a symmetric function).
The whole procedure is applied for all members of 
the ensemble. This extensive averaging is necessary since the patterns 
${\bf \xi}^m$ taken from the long sequences $D_L$, exhibit large fluctuations 
(recall that $N \ll L$ and the variance decreases linearly with the size 
of the sequence).  

%--------------------------
%Place Figure 1 around here
%--------------------------

\begin{figure}[htb]
\centerline{\psfig{figure=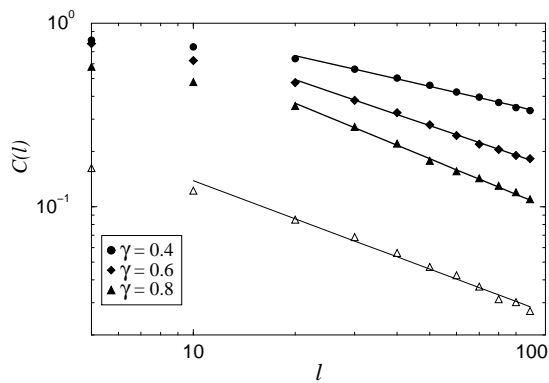,width=7.2cm}}
\caption{The correlation function $C(l)$ (eq.\ \ref{cf_def}) of the sequences 
generated by the trained networks with $N=200$.
The training patterns are generated from 
correlated 1D sequences with $\gamma=0.4,0.6,0.8$. $C(l)$ is shown
along with the power-law regression lines; the respective exponents are 
$0.42, 0.63, 0.76$. The opaque triangle points correspond to training by 
sliding N-sites each cycle, and the exponent of the regression line 
is $0.7$ ~(for $\gamma=0.8$).}
\label{fig_learn_slope}
\end{figure}

Figure \ref{fig_learn_slope} depicts the results of the above procedure for 
$\gamma=0.4,0.6,0.8~~(N=200)$, with $L \approx 10^5$ and an 
ensemble of 50--100 samples. For comparison, we show the results of training
with patterns obtained by sliding the window N-sites each cycle 
(non-overlapping windows, eq.\ \ref{train_set_par}); 
in this case, the sequence is much larger, 
$L \approx 10^7$. The data points are the average 
values with relative error-bars that vary from $5 \%$ for small $l$
to $40 \%$ (less than $20 \%$ for non-overlapping windows)
of the data point for $l \approx N/2$, 
hence, we omit them to preserve the clarity of the figure. 
The learning rate used, $\eta=2$ (eq.\ \ref{weight_update}), is not optimal;
it is obvious that an optimization can reduce the fluctuations in the
correlation function since it affects the relative change in the weights. 
We note that the fluctuations inspected in the sequences $\{\sigma_m\}$
generated by the trained networks are similar to those of the training
patterns ${\bf \xi}$, indicating a finite size effect. 
This has been confirmed for several network sizes.

\begin{figure}[htb]
\centerline{\psfig{figure=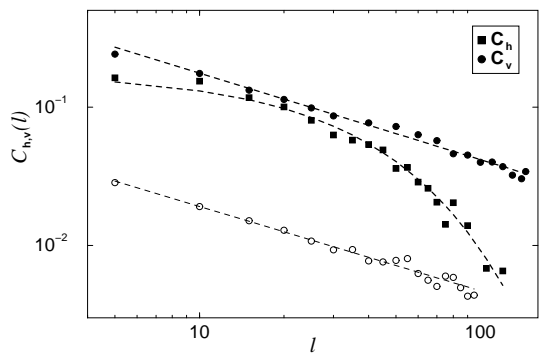,width=7.2cm}}
\caption{The correlation functions of the weights ${\bf W}$ of trained  
networks with $N=300$. $C_{h (v)}$ is averaged over the rows (columns)
of the weight matrix (eq.\ \ref{horver_cor}).  
The dashed lines correspond to 
regression fits: a power-law $~C_v(l) \sim l^{-0.625 \pm 0.016}$ ,
and exponential $~C_h(l) \sim \exp(- a~ l)~$ $a=-0.03\pm0.001$. 
The opaque circles represent $C_v$ in the case of training by sliding N-sites 
each cycle for a network with $N=200$. The power-law regression fit is 
$~C_v(l) \sim l^{-0.61 \pm 0.025 }$ }
\label{fig_net_learn_cor}
\end{figure}

It is interesting to examine how the network (learning algorithm) has embedded
the relevant information associated with the correlations. 
At the end of the training process described above, we measured the
correlation function of the weight vectors (averaged over realizations 
of $D_L$) in two directions, horizontal (over rows) and vertical 
(over columns), as follows:
\beq\label{horver_cor}
\begin{array}{ccll}
C_h(l) & = & \left\langle 
\sum_{i,j=1}^N W_{i,j} ~ W_{i,j+l\bmod N} \right\rangle
 & \mbox{- horizontal} \\
C_v(l) & = & \left\langle 
\sum_{i,j=1}^N W_{i,j} ~ W_{i+l\bmod N,j} \right\rangle
 & \mbox{- vertical}
\end{array}
\eeq
\noindent
Results are presented in Fig.\ \ref{fig_net_learn_cor} for a training sequence
obeying a power-law correlation function with exponent $\gamma=0.6$ and a 
network of size $N=300$.
Clearly, the vertical correlations follow a rule similar to that of the 
sequence, $~C_v(l) \sim l^{-0.625}~$,
while the horizontal correlations decay much faster, with an exponential fit 
$C_h(l)\sim \exp(- 0.03~ l)$. The case of training with patterns obtained from
non-overlapping windows is presented for comparison by the opaque circles. 
In this case, the vertical correlations are similar, 
$~C_v(l) \sim l^{-0.61}~$, however there are no horizontal correlations. 
We conclude that sliding one site each cycle induces horizontal correlations,
however they decay exponentially fast. 

%--------------------------
%Place Figure 2 around here
%--------------------------

The issue of learning the rule from a teacher is not treated in this 
framework, however, we add the following (numerical) observation: 
The overlap between two 
(initially random) networks, $R={\bf W}^1 \cdot{\bf W}^2$,
learning from the same rule (training sequence) decreases with the size 
of the network ($N$) and remains very low even for $L \gg N^2$, although the 
two networks generate sequences with similar correlation functions, 
asymptotically. The same holds when each network learns from a different 
sequence. 
We conclude that the network indeed learns the statistical properties of the
sequence, but not its values. Clearly, in batch learning one expects that the
network would learn the rule when $L=N^2$, since the number of free parameters
 (weights) equals the number of examples. 
 
\section{Generating colored sequences by a colored network}
\label{sec_col_net}

So far, we demonstrated the capability of the network
to capture statistical properties from the training sequence.
Let us now consider the inverse problem, i.e., that of constructing a network
that is capable of generating correlated sequences. 
The information obtained from the trained networks in the previous section 
regarding the structure of the weight matrix, suggests that the significant 
correlation is present between the elements of a column, i.e., 
vertical direction. 
Therefore, we would like to compare the correlation function of 
sequences generated by networks with the same vertical correlation function 
(power-law), and various horizontal decay forms, i.e., power-law with an 
increasing exponent, $\gamma$.  The weights are constructed as follows. 
Start by generating a random matrix of normally distributed elements. 
Each column is treated as a 1D sequence and is ``colored'' following the 
process described above for generating a 1D correlated sequence. 
After this stage, the rows are still uncorrelated. To achieve a different 
power-law function for the rows, we treat each row independently as a 1D 
sequence and follow the same procedure as above, this time with (possibly) 
a different exponent. 
This process generates a weight matrix with pronounced correlations
in the vertical and horizontal directions only. We normalize the weights, 
$~\sum_{i,j=1}^N W_{(i,j)}^2 = N~$, such that $~\beta= {\cal O}(1)$ 
(independent of $N$). The value of $\beta$ in the dynamic equations, 
eqs.\ \ref{updating_rule_seq}-\ref{updating_rule_par}, 
is taken well above bifurcation to increase the 
probability of non-periodic attractors \cite{kan96} 
(we carefully avoid the periodic attractors in our measurements).  
In the analysis described below, each sample network (colored ${\bf W}$) is 
initialized at random (${\bf S}^0$) and the correlation function of
the sequence generated is calculated at long times.
%
%--------------------------
%Place Figure 3 around here
%--------------------------

\begin{figure}[htb]
\centerline{\psfig{figure=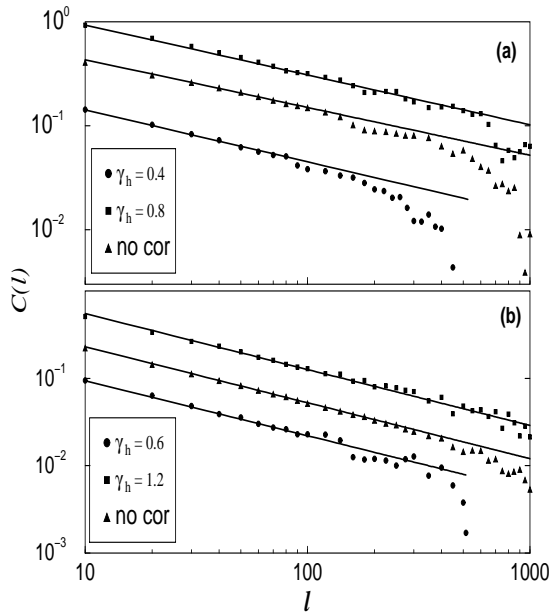,width=7.2cm,height=8.3cm}}
\caption{The correlation function of sequences generated by a constructed 
colored network, $N=2048$. (a) $\gamma_v=0.4$ (b) $\gamma_v=0.6$.
$\gamma_h$ is given in the figure
(``no cor'' stands for no horizontal correlations). 
The solid lines are the power-law 
regression fits with exponents: 
(a) $0.47$ (top line), $~0.46$ (middle) and  $~0.5$ (bottom)
(b) $0.64$ (top), $~0.64$ (middle) and  $~0.63$ (bottom)
}
\label{fig_corgen_Gc4_6}
\end{figure}

Figure \ref{fig_corgen_Gc4_6} depicts two cases for which the vertical 
correlation function of the weights decays polynomially with exponents
$\gamma_v=0.4$ and $\gamma_v=0.6$. For each case, the horizontal
correlation function takes one of the following three values: 
$\gamma_h=\gamma_v$, $~\gamma_h=2 \gamma_v$ or uncorrelated. 
The results were obtained for a network
of size $N=2048$ and averaged over 50 realizations of the weight matrix. 
Additional averaging is done by starting from several initial conditions for
each matrix. 
It is apparent that the symmetric case, $\gamma_h=\gamma_v$, gives rise to 
a relatively poor long-range correlations in the generated sequence. 
The other two cases exhibit much longer-range
correlations. Although we are not trying to determine the optimal correlation
function, it seems that weak correlations are better than lack of horizontal
correlations. This is in agreement with our findings regarding the trained 
networks, see Fig.\ \ref{fig_net_learn_cor}. 

To conclude, we propose a naive calculation which should serve as a starting
point to the analytical investigation of the model. 
The quantity of interest in our calculation is the asymptotic correlation 
function,  
\beq\label{cor_norm}
C(l) = Z ~\left\langle~ S_i^t ~ S_{i+l}^t ~\right\rangle_{t,W} \quad ,
\eeq
\noindent 
where $Z$ is a normalization factor ($C(0)=1$). The average is taken over
the time $t$ and the realizations of the weights, expecting $C(l)$ to be
independent of the site $i$.
For simplicity we use the parallel updating rule, 
eq.\ \ref{updating_rule_par}, 
hence, the stationary state of the correlation function may be given by
\beq\label{cor_sum}
C(l) = Z ~
\left\langle \tanh \left(\beta \sum_{j=1}^{N} S_j^t W_{i,j} \right) \tanh 
\left(\beta \sum_{j=1}^{N} S_j^t W_{i+l,j} \right) \right\rangle_{t,W} .
\eeq
\noindent
The approximation of eq.\ \ref{cor_sum} consists of linearizing the r.h.s.\ 
 and assuming ${\bf S}$ independent of the
realization of ${\bf W}$, leading to 
\beq\label{cor_mf1}
C(l) \approx  \sum_{j,k=1}^{N} 
\left\langle S_j^t S_k^t \right\rangle_t ~ \left\langle W_{i,j} W_{i+l,k}
\right\rangle_W \quad .
\eeq     
\noindent
Next, we identify the averages as the correlation 
functions, defined above, that depend on the distance only, and
rewrite eq.\ \ref{cor_mf1} using $~m \equiv k-j~$ in the form
\beq\label{cor_mf2}
C(l) = {\hat Z} \sum_{m=1}^{N} ~C(m) ~C_W(l,m)  \quad ,
\eeq     
\noindent
where ${\hat Z}$ is a normalization factor, 
$C_W(l,m)$ denotes the 2D correlation function of the weights, 
and $C(l)$ is the 1D function which is the quantity of interest. 
In the scenario described above, $C_W$ is known a priori (independent
of time) since the weights are constructed. 
If we assume a power-law vertical correlations for $C_W$ and no horizontal
correlations, the correlation function of the sequence, $C(l)$, 
simply follows the vertical correlations of the weights, i.e.,
$~C(l)= {\hat Z} \sum_m C_W(l,m) \delta_{m,0} = {\hat Z} C_W(l,0)~$,
supporting the above findings. We remark that 
this decomposition of $C_W$ into independent vertical and horizontal functions
is still a good approximation when $C_h$ is not a delta function, 
as long as $C_h$ decays much faster than $C_v$, which enables us to neglect 
other correlations. In this case eq.\ \ref{cor_mf2} can be formulated as an 
``Eigen-value problem'' of the matrix $C_W$. Few such cases have been solved
numerically for which the assumption of decomposition was found consistent, 
see \cite{priel_thes}.
When this decomposition is no longer valid, we observe a breakdown of the
long-range power-law behavior, see Fig.\ \ref{fig_corgen_Gc4_6} - the case
$\gamma_h=\gamma_v$.  

\section{Discussion}

In this paper we analyzed the capability of a neural network model to learn
the rule of a long-range correlated sequence on the one hand, and a method 
for constructing a network that is able to generate such sequences on the 
other hand. 
We demonstrated that a simple on-line learning algorithm can be used to 
extract the rule, provided that the sequence is long enough. 
The fluctuations observed in the training patterns are manifested in the 
generated sequences as well. The investigation of the weights that were 
obtained during the learning process indicates that the vertical correlations
($C_v$, eq.\ \ref{horver_cor}) play the most important role in generating 
correlated sequences by the network. Employing this finding, we were able to
construct networks (without training) that are capable of generating sequences
with a predefined power-law correlation function. Indeed, we found that 
additional significant horizontal correlations in the constructed networks'
weights corrupt this property. These observations were confirmed by a naive
analytical treatment of the stationary correlation function.

The question of the optimal learning rate ($\eta$) was not treated although 
it seems to be an important parameter in the convergence of the training 
process. Another question which deserves further research regards the 
analytical
derivation of the correlation function. Taking into account the nonlinearity 
of the transfer function is necessary to close the naive calculation 
self-consistently, and to obtain the corrections to the correlation function.

\section*{Acknowledgments}
We thank W. Kinzel for fruitful discussions and critical reading of the 
manuscript. IK acknowledges the support of the Israel Ministry of Science.

%----------------------------------------------------------------

\end{document}